# Machine Learning Methods for Network Intrusion Detection

Mouhammad Alkasassbeh, Mohammad Almseidin

*Abstract*—Network security engineers work to keep services available all the time by handling intruder attacks. Intrusion Detection System (IDS) is one of the obtainable mechanisms that is used to sense and classify any abnormal actions. Therefore, the IDS must be always up to date with the latest intruder attacks signatures to preserve confidentiality, integrity, and availability of the services. The speed of the IDS is a very important issue as well learning the new attacks. This research work illustrates how the Knowledge Discovery and Data Mining (or Knowledge Discovery in Databases) KDD dataset is very handy for testing and evaluating different Machine Learning Techniques. It mainly focuses on the KDD preprocess part in order to prepare a decent and fair experimental data set. The J48, MLP, and Bayes Network classifiers have been chosen for this study. It has been proven that the J48 classifier has achieved the highest accuracy rate for detecting and classifying all KDD dataset attacks, which are of type DOS, R2L, U2R, and PROBE.

## I. Introduction

BUILDING a reliable network is a very difficult task considering all different possible types of attacks. Nowadays, computer networks and their services are widely used in industry, business, and all arenas of life. Security personnel and everyone who has a responsibility for providing protection for a network and its users, have serious concerns about intruder attacks.

Network administrators and security officers try to provide a protected environment for users' accounts, network resources, personal files and passwords. Attackers may behave in two ways to carry out their attacks on networks; one of these ways is to make a network service unavailable for users or violating personal information. Denial of service (DoS) is one of the most frequent cases representing attacks on network resources and making network services unavailable for their users. There are many types of DoS attacks, and every type has it is own behavior on consuming network resources to achieve the intruder's aim, which is to render the network unavailable for its users [1]. Remote to user (R2L) is one type of computer network attacks, in which an intruder sends set of packets to another computer or server over a network where he/she does not have permission to access as a local user. User to root attacks (U2R) is a second type of attack where the intruder tries to access the network resources as a normal user, and after several attempts, the intruder becomes as a full access user [2]. Probing is a third type of attack in which the intruder scans network devices to determine weakness in topology design or some opened ports and then use them in the future for illegal access to personal information. There are many examples that represent probing over a network, such as nmap, portsweep, ipsweep.

IDS becomes an essential part for building computer network to capture these kinds of attacks in early stages, because IDS works against all intruder attacks. IDS uses classification techniques to make decision about every packet pass through the network whether it is a normal packet or an attack (i.e. DOS, U2R, R2L, PROBE) packet.

KDD is an online repository dataset, which includes all types of intruders' attacks such as DOS, R2L, U2R, and PROBE. In this paper, a number of classifiers will be evaluated on the KDD dataset. The methodology followed in this study is first to perform a preprocessing step on KDD dataset and after to use the prepared dataset on a fair environment and resources, and finally, to examine which classifier is more accurate than others in detecting all studied attacks (DOS, R2L, U2R, and PROBE).

The remainder of this work is organized as follows; related work is presented in Section II, which also provides brief discussion about KDD dataset and selected classifiers. Section III gives detailed steps of the preprocessing approach performed on the KDD dataset. The used classification techniques are explained in Section IV. Experiments and classifiers evaluation are presented in Section V. Section VI presents a comprehensive comparison between the selected classifiers and experimental results with statistical values, followed by conclusions and future work in Section VII.

## II. Related Work

IDS combines hardware and software to detect attacks on networks in order to ensure the protection of the system from unauthorized access. IDS can be divided into two main classifications based on the attack's detection method. The first one is the misuse, and the second is anomaly detection. The anomaly detection can be used in different ways in order to detect any strange behavior of the user within the network traffic.

IDS built on Artificial Neural Network (ANN) and fuzzy clustering (FC) has been proposed to find out some networks problems and attacks. However, there are limitations of this proposed model; for example, it has a lack of accuracy in low-frequent attacks. The researchers here took over this limitation by dividing heterogeneous training set into homogeneous

M. Alkasassbeh is with Mutah University, Jordan (e-mail: mouhammd.alkasassbeh@mutah.edu.jo).

M. Almseidin is with University of Miskolc, H-3515 Miskolc, Hungary (e-mail: alsaudi@iit.uni-miskolc.hu).

training subsets. By reducing the complexity of each sub-training set, the performance of detection is increased, and the backup of the system can be taken successfully by using restore point [3].

Artificial intelligence techniques with heuristic algorithms such as Genetic Algorithm (GA) and ANN are used in IDS gaining its ability to learning and development, which makes them more accurate and efficient in facing the increasing number of unpredictable attacks. GA and ANN combined approach gives the IDS with extra performance and accuracy [4].

In the work of Pradhan et al. [5], they took into account the user actions as a parameter in anomaly detection using a back propagation in their test. Their work is very promising. The back propagation neural network had a classification rate of 100%. The detection rate was 88% on attacks in general whether known or unknown attacks. The main advantage of this work is the minimum amount of training data that needs to give good results of classification the traffic.

Recently, an improvement alternative of ANN is proposed called Multi-Layer Perception (MLP) ANN. The MLP method made ANN IDS methods more accurate and efficient in terms of detection and normal communication. The MLP-ANN method shows detection result much better than traditional methods. MLP overcomes the limitation of detection low-frequency attacks. In addition, MLP-ANN IDS can define the type of attacks and classify them. This feature allows system to predefine actions against similar future attacks [6]- [8].

In the classifier selection model presented by Nguyen and Choi [9], they extracted 49,596 instances of KDD dataset and compared a set of classifiers under control environment. Lahreet et al. [10] presented different approaches to deal with KDD dataset, supervised, and unsupervised methods simulated using MATLAB, and researchers test supervised and unsupervised techniques with fuzzy rules to identify the performance of the proposed system. Breiman [11] focused on random forest and how is it combined between trees predictors, and the researcher proposed error in the random forest as limit number of trees in the forest.

Bhargava et al. [12] illustrated in decision tree analysis on j48 algorithm and how it is important to calculate entropy and information gain for each attributes in any dataset ready to be classified, they used decision tree with univariate and multivariate methods, also the researchers presented multivariate method as linear machine method. The researchers recommended this approach for large amount of data.

Fleizachet et al. [13] stated that nature of dataset sometimes makes it difficult to select appropriate attributes to learn, and the researchers implement experiments with Naïve Bayes classifier and measure performance for each call.

III. KDD PREPROCESSING

MIT Lincoln labs provided KDD dataset [14], it is very helpful to examine which classifier demonstrates high accuracy to detect (DOS, R2L, U2R, and PROBE) attacks. In our work, the KDD dataset has imported to Oracle database server, because there was a need to extract fairly experimental dataset for a set of classifiers with statistical information about each type of attack at KDD dataset, also to collect statistical information about each attack type instance. Table I illustrates KDD dataset after importing it to the database server, and the table also lists number of instances for each type of attack.

TABLE I
NUMBER OF INSTANCES FOR EACH TYPE OF ATTACK

| Attack Type | Number of instances |
| --- | --- |
| SMURF(DOS) | 2,807,886 |
| NEPTUNE(DOS) | 1,072,017 |
| Back (DOS) | 2,203 |
| POD (DOS) | 264 |
| Teardrop (DOS) | 979 |
| Buffer overflow (U2R) | 30 |
| Load Module (U2R) | 9 |
| PERL (U2R) | 3 |
| Rootkit (U2R) | 10 |
| FTP Write (R2L) | 8 |
| Guess Passwd (R2L) | 53 |
| IMAP(R2L) | 12 |
| multihop (R2L) | 7 |
| PHF (R2L) | 4 |
| SPY (R2L) | 2 |
| Warez client (R2L) | 1,020 |
| Warez Master (R2L) | 20 |
| IPSWEEP (PROBE) | 12,481 |
| NMAP (PROBE) | 2,316 |
| PORTSWEEP(PROBE) | 10,413 |
| SATAN (PROBE) | 15,892 |
| Normal | 972,781 |

TABLE II
NUMBER OF INSTANCES AFTER ORGANIZATION

| Attack Type | Number of instances |
| --- | --- |
| SMURF(DOS) | 85,983 |
| NEPTUNE(DOS) | 32,827 |
| Back (DOS) | 70 |
| POD (DOS) | 10 |
| Teardrop (DOS) | 30 |
| Buffer overflow (U2R) | 10 |
| Load Module (U2R) | 2 |
| PERL (U2R) | 1 |
| Rootkit (U2R) | 5 |
| FTP Write (R2L) | 2 |
| Guess Passwd (R2L) | 10 |
| IMAP(R2L) | 4 |
| multihop (R2L) | 2 |
| PHF (R2L) | 1 |
| SPY (R2L) | 1 |
| Warez client (R2L) | 31 |
| Warez Master (R2L) | 7 |
| IPSWEEP (PROBE) | 382 |
| NMAP (PROBE) | 70 |
| PORTSWEEP(PROBE) | 318 |
| SATAN (PROBE) | 487 |
| Normal | 28,500 |

We have 21 types of attacks, categorized into four main groups with different number of instances and occurrences. After cleaning and removing the duplicated records, we extract a new KDD dataset, all instances of experiment are fully randomized; we have the following table with 148,758 instances organized as follows (Table II):

After preparing the KDD dataset for classification experiment techniques, the idea for the next step is to work with the most common used classifier: multilayer perception, Bayesian algorithm, trees and rules using Waikato Environment for Knowledge Analysis (WEKA) software.

## IV. CLASSIFICATION TECHNIQUES

### A. J48 Tree

This was first introduced by [15]. It is the most common classifier used to manage the database for supervised learning that gives a prediction about new unlabeled data, J48 creates Univariate Decision Trees. J48 based used attribute correlation based on entropy and information gain for each attributes [12]. It has been used in many fields of study, such as data mining, machine learning, information extraction, pattern recognition, and text mining. It has many advantages; it is capable of dealing with different input data types: numeric, textual and nominal. J48 decision tree is an extension of the algorithm ID3. It has an advantage over ID3 in that it can build small trees. It follows a depth-first strategy, and a divide-and-conquer approach.

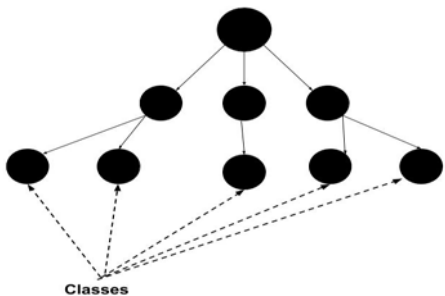

Fig. 1 Decision Tree Structure

A decision tree consists of several elements: root, internal nodes, and leaves. The internal nodes represent the conditions in which the value of the parameters will be tested. Based on these values and the condition, the flow of the tree will be decided (along which branch the decision tree must go). Leaf nodes represent the decision or the class. Fig. 1 shows a typical decision tree structure.

The tree is constructed by following these three main steps:
1. Ensure that all of the grouped inputs are of the same class. Then ensure that the tree is labelled with the class.
2. Calculate some parameters for each attribute, such as information gain.
3. Choose the best split attribute based on the criteria that have been set.

Entropy comes from information theory; it indicates the amount of information that is held; in the other words, the higher the entropy, the more information content there is. It can be measured by:

$$\text{Entropy} = \sum_i -p_i \log_2 p_i \quad (1)$$

where Pi is the probability of the class 'i'.

Information gain expresses the importance of the feature or attribute, and it determines which attribute is the most important one for distinguishing between the classes to be knowledgeable. This piece of information is calculated also on training data. Information gain can help in choosing the best split; if it has a high value then this split is good, otherwise the split is not good enough. Information gain can be calculated by the data achieved from entropy:

Information Gain = entropy (parent) – [average entropy (children)] (2)

### B. Multilayer Perceptron (MLP)

MLP is widely used neural network classifier based on number of classes (output) and number of hidden layers, MLP uses weights for every node at neural network, most effective attributes will get large weights conversely attributes not affect in predictive class. MLP always takes largest time for training, but it has quick time for testing [16]. MLP has been positively used in daily life uses like regression problems, classification and prediction problems.

An example of a modest structure MLP network is illuminated in Fig. 2. MLP drives the data flow to be taken in one direction from input to output. As there will be no feedback, According to [17] and [18], any MLP network can be notable by a number of performance features, which can be brief in three points:
1. Neural Network Architecture: Overall, MLP architecture can be clarified as set of links between the neurons in different layers. Generally, the architecture consists of three main layers: input layer, hidden layers and output layer. MLP is most of the time fully connected. On each link there is a weight, which is tuned based on the training algorithm.
2. Training Algorithm: is the method of selecting one model from a set of models, which tunes the weights of the links.
3. Transfer Function: is applied on the net input of each neuron to control the net output signal. Here in, the function is usually non-linear. The most common function used as transfer function is Sigmoid function. The use of the sigmoid function has an advantage in neural networks trained by a back propagation learning algorithm. Table III illustrates examples of some common transfer functions.

To understand how the learning process on MLP is done, here is a simple example to demonstrate the process, suppose that we have an MLP, which has N neurons as input layer and M neurons in the hidden layers, and single output neuron. The learning process will be as follows:
1. Hidden layer stage: Given a number of inputs I (the output of the input layer) and a set of equivalent weights as also an input between the input and hidden neurons wij,

then the outputs of all neurons in the hidden layer are calculated as in (3) and (4):

TABLE III
TRANSFER FUNCTIONS [17]

| Transfer function | Definition |
|---|---|
| Linear | $f(x) = x$ |
| Sigmoid | $f(x) = \frac{1}{1+e^{-x}}$ |
| Hyperbolic | $f(x) = \frac{e^x - e^{-x}}{1+e^{-x}}$ |
| Hard limit | $f(x) = \begin{cases} 0, x < 0 \\ 1, x \geq 0 \end{cases}$ |
| Symmetric hard limit | $f(x) = \begin{cases} -1, x < 0 \\ 1, x \geq 0 \end{cases}$ |

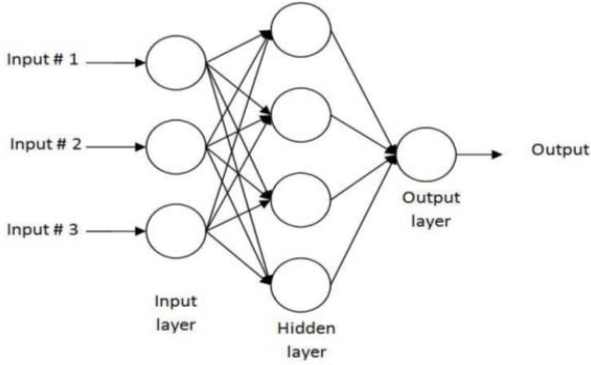

Fig. 2 MPL architecture

$$O_i = \sum_{i=0}^{N} w_{ij} \psi_i \quad (3)$$

$$y_j = z(O_j) \quad (4)$$

where i = 1, 2 ..., N and j = 1, 2... M. The z and $yj$ are the activation function and output of the $j^{th}$ node in the hidden layer, respectively. The z is usually a sigmoid function which given in (5).

$$z(x) = \frac{1}{1+e^{-x}} \quad (5)$$

2. Output stage: Equation (6) is the final outputs of all neurons in the output layer. For simplicity, the equation below explains the output:

$$Y^\wedge = f(\sum_{J=0}^{m} wJ \, y_J^H) \quad (6)$$

where *f()* is the activation function of the output layer, which is typically a linear function. And the $Y^\wedge$ is the output of the neural network. The MLP network is always trying to make the error very small through the Back Propagation (BP) Training algorithm. At the beginning, all the weights are initialized with a random value, and after that, the weights are changing in each iteration until satisfied state values are obtained.

3. Error validation stage: ANN keeps learning until the error becomes very small assuming that the observed output is Y and the predicted output is ˆY. The learning process will keep going until the error difference given in (7) is a minimum value, as the minimum is the best. N is the total number of instances that used during training stage.

$$Error = \frac{1}{N}\sum_{i=1}^{T}(Y_i - \hat{Y}_i)^2 \quad (7)$$

In MLP, the weights and bias values are allocated randomly, here in, the goal of the training is to find the set of weights that give the output of the network to be close as possible to the real values.

*C. Bayes Network*

It is a classifier for supervised learning that uses assumptions of independent features. It uses theory of learning that represents distribution naïve Bayesian classifier. It uses various search algorithms and different quality measure methods. Bayes Network is an enhancement for Naïve Bayes [19].

A Bayesian network is very useful, because it helps us to understand the world that we are modeling. BayesNet may be the best in various areas of life, where modeling a mysterious fact and in the state of decision nets, wherever it is good to make intelligent, justifiable and quantifiable decisions that will enhance performance of classification. In brief, BayesNet is helpful for diagnosis, prediction, modeling, monitoring and classification [20].

The main idea of the Bayesian classifier consists of two phases: in the first, if an agent has an idea and knows the class, in this case it can predict the values of the other features; in the second, if the agent does not have an idea or does not know the class, in this case the Bayes rule is used to predict the class given.

We used the Bayesian Network as a classifier for the following reasons:
- Probabilistic learning, which calculates clear probabilities for assumption.
- Incremental, which is a prior knowledge and possible to be added to data viewing.
- Probabilistic prediction, which can predict more than one hypothesis, weighted by the probabilities.

The theory of the Bayesian Network is shown in (8), where the symbol D indicates the training data, the probability of hypothesis h.

$$P(h|D) = \frac{P(D|h)P(h)}{P(D)} \quad (8)$$

The symbols in (8) refer to:
- P (h|D): posterior probability.
- P (D|h): condition probability.
- P (h): prior probability of h.
- P (D): marginal probability of D.

V. PERFORMANCE EVALUATION OF THE SELECTED CLASSIFIERS

KDD dataset presents real packets focused on wired network; it has 41 features about each packet that will help to implement different classifier types. The current experiments that are performed present fair test environment because we extracted 148,758 instances from all four groups of attack

(DOS, R2L, U2R, and PROBE) as training dataset, normal packets present %19 from current experiment as original KDD dataset normal packets and the highest proportion for the DOS attack with 79% from current experiment as original KDD dataset DOS packets.

For fair control comparison between different classifiers, another 60,000 independent instances were extracted from original KDD dataset as test sample and these instances fully randomized and not included in training dataset. The experiment environment applied with Weka version 3.7.12 and Intel Xeon (R) CPU E5-2680 @ 2.70GHzX4 with available RAM 8.0GB under Ubuntu 13.10 platform. Most common classifiers are used in this experiment (J48, Random forest, Random Tree, Decision Table, Multilayer Perceptron (MLP), Naïve Bayes and Bayes Network). All models and results are saved to start comprehensive study about which classifier has the highest accuracy rate to detect attacks.

## VI. EXPERIMENT EVALUATION AND RESULTS

All selected classifiers tested with 60,000 independent instances from KDD dataset and all test instances are fully randomized. This section illustrates all parameters values that have been used in selected classifiers in the experiments.

Table IV lists statistical values that achieved in our experiments and it can be seen that random forest classifier achieves the highest Kappa statistic with rate equal to 0.8957 and the lowest Kappa statistic with Bayes network classifier with rate equal to 0.8464.

Table V records weighted average for true positive (TP) and false positive (FP) for each classifier selected for experiment, the J48 achieves the highest TP rate with value equals to 0.931.

Table VI presents accuracy rate that recorded in the experiment. Also, The J48 classifier achieves the highest rate accuracy.

TABLE IV
STATISTICAL VALUES

| Classifier | Kappa statistic | Mean absolute error | Root mean squared error |
|---|---|---|---|
| J48 | **0.8844** | **0.0059** | **0.0763** |
| MLP | 0.8639 | 0.0075 | 0.0813 |
| Bayes Network | 0.8464 | 0.0085 | 0.087 |

TABLE V
WEIGHTED AVERAGE FOR TRUE POSITIVE (TP) AND FALSE POSITIVE (FP)

| Classifier | TP Rate | FP Rate | Precision | ROC Area |
|---|---|---|---|---|
| J48 | **0.931** | 0.005 | 0.989 | 0.969 |
| MLP | 0.919 | 0.014 | 0.978 | 0.990 |
| Bayes Network | 0.907 | 0.000 | 0.992 | 0.999 |

TABLE VI
ACCURACY RATE

| Classifier | Correctly classified Instances | incorrectly classified Instances | Accuracy |
|---|---|---|---|
| J48 | 55865 | 4135 | **93.1083 %** |
| MLP | 55141 | 4859 | 91.9017 % |
| Bayes Network | 54439 | 5561 | 90.7317 % |

## VII. CONCLUSIONS AND FUTURE WORK

Due to the urgent demand for an effective IDS in network security, researchers are striving to identify improved approaches. This work illustrates how the KDD dataset is very useful for testing different classifiers. The work concentrates on KDD preprocess phase to prepare fair experiments and fully randomized independent test data. Among the classification techniques (J48, MLP and Bayes Network), the J48 classifier has achieved the highest accuracy rate for detecting and classifying all KDD dataset attack types (DOS, R2L, U2R, and PROBE). KDD dataset has 41 attributes and all of them have been recorded, but as part of future work more classifiers will be tested as well as the feature selection to see the most important features.

REFERENCES

[1] N. Huy and C. Deokjai, "Application of data mining to network intrusion detection: classifier selection model," Challenges for Next Generation Network Operations and Service Management, pp. 399--408, 2008.
[2] S. Paliwal and R. Gupta, "Denial-of-service, probing \& remote to user (R2L) attack detection using genetic algorithm," International Journal of Computer Applications, vol. 60, no. 19, pp. 57--62, 2012.
[3] D. Gaikwad, S. Jagtap, K. Thakare and V. Budhawant, "Anomaly based intrusion detection system using artificial neural network and fuzzy clustering," International Journal of Engineering}, vol. 1, no. 9, 2012.
[4] V. Bapuji, R. N. Kumar, A. Goverdan and S. Sharma, "Soft computing and artificial intelligence techniques for intrusion detection system," Networks and Complex Systems, vol. 2, no. 4, 2012.
[5] M. Pradhan, S. K. Pradhan and S. K. Sahu, "Anomaly detection using artificial neural network," International Journal of Engineering Sciences \& Emerging Technologies, vol. 2, no. 1, pp. 29--36, 2012.
[6] M. Sammany, M. Sharawi, M. El-Beltagy and I. Saroit, "Artificial neural networks architecture for intrusion detection systems and classification of attacks," in The 5th international conference INFO2007, 2007.
[7] M. Al-Kasassbeh, G. Al-Naymat, A. Hassanat and M. Almseidin, "Detecting Distributed Denial of Service Attacks Using Data Mining Techniques," International Journal of Advanced Computer Science and Applications, vol. 7, no. 1, 2016.
[8] M. Al-kasassbeh, "An Empirical Evaluation For The Intrusion Detection Features Based On Machine Learning And Feature Selection Methods," Journal of Theoretical and Applied Information Technology, vol. 22, p. 95, 2017.
[9] H. Nguyen and D. Choi, "Application of data mining to network intrusion detection: classifier selection model," Challenges for Next Generation Network Operations and Service Management, pp. 399--408, 2008.
[10] M. K. Lahre, M. T. Dhar, D. Suresh, K. Kashyap and P. Agrawal, "Analyze different approaches for IDS using KDD 99 data set," International Journal on Recent and Innovation Trends in Computing and Communication, vol. 1, no. 8, pp. 645--651, 2013.
[11] L. Breiman, "Random forests," Machine learning, vol. 45, no. 1, pp. 5--32, 2001.
[12] N. Bhargava, G. Sharma, R. Bhargava and M. Mathuria, "Decision tree analysis on j48 algorithm for data mining," Proceedings of International Journal of Advanced Research in Computer Science and Software Engineering, vol. 3, no. 6, 2013.
[13] C. Fleizach and S. Fukushima, A naive Bayes classifier on 1998 KDD Cup, echnical Report, Department of Computer Science and Engineering, University of California, San Diego, 1998.
[14] D. I. D. DataSet, "Lincoln Labrototy MIT," MIT, (Online). Available: https://www.ll.mit.edu/ideval/data/. (Accessed 5 4 2018).
[15] Breiman, Leo, Friedman, J. H, Olshen, R. A, Stone and C. J, "Classification and regression trees. Wadsworth," Belmont, CA, 1984.
[16] S. K. Pal and S. Mitra, "Multilayer perceptron, fuzzy sets, and classification," IEEE Transactions on neural networks, vol. 3, no. 5, pp. 683--697, 1992.
[17] K. Gurney, An introduction to neural networks, CRC press, 1997.
[18] L. Fausett and L. Fausett, Fundamentals of neural networks:

architectures, algorithms, and applications, Prentice-Hall, 1994.
[19] N. Friedman, D. Geiger and M. Goldszmidt, "Bayesian network classifiers," Machine learning}, vol. 29, no. 3, pp. 131--163, 1997.
[20] S. B. Kotsiantis, I. Zaharakis and P. Pintelas, Supervised machine learning: A review of classification techniques, 2007.